\documentclass[prb,floatfix,twocolumn,superscriptaddress,showpacs,amsmath,amssymb,showpacs]{revtex4-2}

\usepackage{graphicx}
\usepackage{epstopdf}
\usepackage{epsfig}
\usepackage{epsf}
\usepackage{url}
\usepackage[USenglish]{babel}
\usepackage{hyperref}
\def\bcen{\begin{center}}
\def\ecen{\end{center}}
\renewcommand\[{\begin{equation}}
\renewcommand\]{\end{equation}}
\usepackage{verbatim}
\usepackage{xcolor}
\usepackage{natbib}
\usepackage{amsmath, nccmath}
\usepackage{dcolumn}
\usepackage{bm}
\usepackage{bbm}
\usepackage{lipsum}

\begin{document}
\title{Strong coupling impurity solver based on quantics tensor cross interpolation}
\author{Aaram J. Kim}
\affiliation{Department of Physics and Chemistry, DGIST, Daegu 42988, Korea}
\author{Philipp Werner}
\affiliation{Department of Physics, University of Fribourg, 1700 Fribourg, Switzerland}

\begin{abstract}
Numerical methods capable of handling nonequilibrium impurity models are essential for the study of transport problems and the solution of the nonequilibrium dynamical mean field theory (DMFT) equations. In the strong correlation regime, the self-consistently resummed hybridization expansion is an appealing strategy, which however has been employed so far mainly in the lowest-order noncrossing approximation. At higher orders, standard implementations become numerically costly, but a 
significant speed-up can be achieved by evaluating multidimensional integrals in an approximate factorized form. Here we develop a one-crossing approximation solver based on the recently introduced quantics tensor cross interpolation, and demonstrate its accuracy and efficiency with applications to the Anderson impurity model and nonequilibrium steady-state DMFT calculations for the Hubbard model.  
\end{abstract}

\maketitle

\section{Introduction}

Impurity models represent magnetic atoms in a metallic host or quantum dot structures \cite{Anderson1961}. They also play an important role as auxiliary systems, whose solution (augmented by a self-consistency condition) provides the dynamical mean field theory (DMFT) description of lattice models \cite{Georges1996}. For equilibrium calculations, several powerful numerical methods enable an accurate treatment of these models in a wide parameter range, even for multiorbital impurities \cite{Weichselbaum2007,Bulla2008,Gull2011,Bauernfeind2017,Lu2019,Kim2022}. Most of these methods can however not be straightforwardly adapted to nonequilibrium settings, because of sign problems, rapid entanglement growth, truncation errors, or other numerical challenges \cite{Werner2009,Gobert2005,Anders2005}. While significant progress has been made in recent years, both with Monte Carlo \cite{Cohen2014} and diagrammatic \cite{Fernandez2022} approaches, we still lack an efficient yet accurate nonequilibrium impurity solver which could be used routinely in nonequilibrium DMFT studies. 

Most of the existing nonequilibrium DMFT investigations in the strong coupling regime are based on the noncrossing approximation (NCA) \cite{Keiter1971}, which is the lowest order approximation of the self-consistently resummed hybridization expansion. The latter approach is appealing, because at any given expansion order, it yields a conserving approximation and is free from stochastic noise. Most importantly, it has been shown to rapidly converge (with increasing order) towards the exact result in typical nonequilibrium setups \cite{Eckstein2010}. In the case of a single-orbital impurity model, the second order expansion (one-crossing approximation, OCA \cite{Pruschke1989}) gives important corrections to the NCA in the intermediate coupling regime, while the third order approximation (TOA) is essentially exact for effectively hot nonequilibrium states. 

Standard implementations of these higher order schemes become numerically costly, because each order adds two integrations over the Keldysh time contour \cite{Aoki2014} and the number of diagram topologies rapidly increases. If time is discretized into uniform steps and the integrals are evaluated with some quadrature rule, the numerical cost for a given diagram scales naively as $t_\text{max}^{2n}$ with the length of the time interval $t_\text{max}$ and the expansion order $n$. (Some improvement to this scaling can be achieved by factoring out vertex parts \cite{Eckstein2010}.)   

A recent innovation, which promises to overcome this unfavorable scaling, is the tensor cross interpolation (TCI) \cite{Oseledets2010} of multivariable functions defined on such discretized grids. This interpolation allows to express multidimensional integrals as products of one-dimensional matrix integrals~\cite{Oseledets2010,Savostyanov2011,Savostyanov2014}, which is beneficial if the multivariable function can be accurately represented in a compressed factorized form. In  weak-coupling diagrammatic calculations, oscillating integrals up to dimension 40 were successfully evaluated with this technique \cite{Fernandez2022,Matsuura2025}. A second relevant innovation is the quantics representation of multivariable functions \cite{Shinaoka2023}, which enables a compact tensor train representation of functions on an exponentially fine grid by introducing a binary encoding of the variables. The combination of the two techniques, the quantics tensor cross interpolation (QTCI) \cite{Ritter2024}, leverages the power of both methods to provide highly compressed representations of multivariable functions in a factorized form, using an algorithm which samples only a sparse subset of the full function domain. 

Here we implement and test a real-time strong-coupling impurity solver based on QTCI, focusing for simplicity on steady-state problems and OCA. The paper is organized as follows. In Sec.~\ref{sec:method}, we briefly introduce the pseudoparticle formalism and explain how the pseudoparticle self-energy can be evaluated using QTCI. In Sec.~\ref{sec:results}, we test the accuracy of the QTCI implementation for OCA, and show some applications to steady-state DMFT. Section~\ref{sec:conclusions} contains the conclusions.

\section{Method}
\label{sec:method}

\subsection{Pseudoparticle formalism}
\label{sec_pp}

The self-consistently resummed hybridization expansion on the three-leg Kadanoff-Baym contour has been discussed in detail in Refs.~\cite{Eckstein2010,Aoki2014} and aspects relevant to the steady-state implementation on the two-branch Keldysh contour were presented in Ref.~\cite{Li2021}. 
This method is based on the mapping between the local states, e.g. $|0\rangle, |\!\!\uparrow\rangle, |\!\!\downarrow\rangle, |\!\!\uparrow\downarrow\rangle$ for the Anderson impurity model  (AIM) \cite{Anderson1961}, and the pseudoparticle (pp) states generated by adding one corresponding pp to the pp vacuum state.
The resulting pp impurity action becomes noninteracting in the absence of the bath, and the hybridization term plays the role of a retarded interaction between pp's \cite{Keiter1971,Coleman1984,Bickers1987}.

The pp mapping however generates unphysical states which do not exist in the original impurity model, such as the zero pp state, two pp states, and so on.
One can project out such unphysical states at the level of the Feynman diagrams by eliminating 
the pp Green's function lines which propagate backward in time~\cite{Keiter1971,Coleman1984,Bickers1987,Eckstein2010}.
This restriction, which also implies the absence of pp loops, is the main difference to conventional weak-coupling diagrams.

\begin{figure}[t] 
\begin{centering}
\includegraphics[width=0.90\columnwidth]{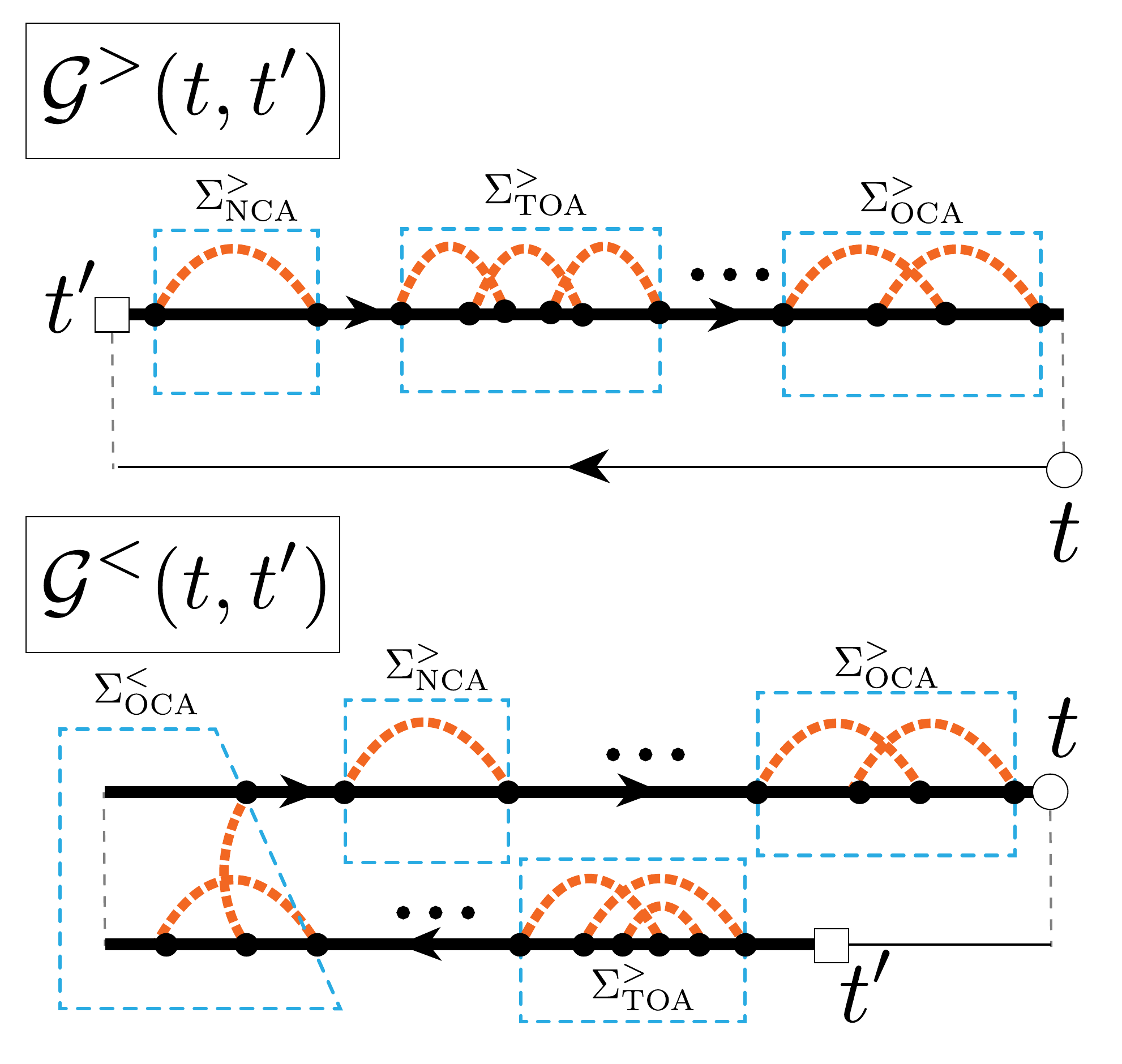}
\par\end{centering}
\caption{
Schematic examples of pp greater and lesser Green's functions constructed via the pp Dyson equation. The square at time $t'$ represents a creation operator, the circle at $t$ an annihilation operator, thick black lines are pp propagators, orange dashed lines hybridization functions and the self-energy insertions are marked by the blue dashed lines. 
\label{Fig:pseudoparticles}
}
\end{figure}

The interacting pp Green's function $\mathcal{G}$ is obtained by solving a pp Dyson equation in which the pp self-energy $\Sigma$ at order $n$ contains all one-particle irreducible diagrams formed by a backbone of pp Green's functions and $n$ hybridization lines (Fig.~\ref{Fig:pseudoparticles}). For $n=1$ this scheme is called noncrossing approximation \cite{Keiter1971}, because the pp Dyson equation sums up all diagrams of the bare hybridization expansion which do not contain any crossings of hybridization lines. Order $n=2$ corresponds to the one-crossing approximation \cite{Pruschke1989}, since the self-energy insertions contain at most two hybridization lines (with one crossing), order $n=3$ includes the self-energy insertions with up to three hybridization lines (third-order approximation, TOA), etc. 

In the case of the single-orbital AIM, the local Hilbert space has dimension $4$ and the pp Green's functions and self-energies can be represented as $4\times 4$ matrices. Disregarding the orientations (forward/backward) and spin flavors of the hybridization lines, there are $(1, 1, 4, \ldots)$ different irreducible topologies of the pp self-energy diagrams at order ($n=1$, $2$, $3$, \ldots). For spin-diagonal hybridization functions and a spin-nonconserving local Hamiltonian $H_\text{loc}$, those topologies include $(4, 16, 256, \ldots)$ self-energy diagrams. 

As mentioned above, the pp Green's functions always propagate in the forward direction along the Keldysh or Kadanoff-Baym contour \cite{Aoki2014}.
This implies that both the retarded (R) and advanced (A) Green's functions can be expressed solely by the greater ($>$) component \cite{Li2021} 
\begin{align}
	\mathcal{G}^R(t,t') &= \theta(t-t')\mathcal{G}^>(t,t')~,\\
	\mathcal{G}^A(t,t') &= -\theta(t'-t)\mathcal{G}^>(t,t')~,
	\label{<+label+>}
\end{align}
where $\theta(t-t')$ is the Heaviside step function. On the Kadanoff-Baym contour with Matsubara branch $\left(0,-i\beta\right)$, one can define the lesser ($<$) component of the pp Green's function in terms of the right-mixing ($\ulcorner$) and greater components,
\begin{equation}
	\mathcal{G}^<(t,t') = -i\xi \mathcal{G}^>(t,0^+)\mathcal{G}^{\ulcorner}(-i\beta,t')~,
	\label{<+label+>}
\end{equation}
where $\xi$ is a diagonal matrix whose $m$-th element is $+1$ ($-1$) if the $m$-th pp state is associated with an even (odd) number of electrons.  

For the description of steady states, whose memory of the initial equilibrium condition is washed out, the Matsubara branch can be pushed to $t\rightarrow -\infty$ and decoupled from the Keldysh branch. 
Then, it is sufficient to consider the greater and lesser components of the pp Green's functions. Since these functions depend only on the relative time $t-t'$ in the steady state, we can work either in a time or frequency representation. In frequency space, the Dyson equations for the retarded and lesser components become \cite{Li2021}
\begin{align}
	\mathcal{G}^{R}(\omega) &= \left[\omega-H_\text{loc} - \Sigma^R(\omega)\right]^{-1}~,\\
        \mathcal{G}^{<}(\omega) &= \mathcal{G}^{R}(\omega)\Sigma^{<}(\omega)\mathcal{G}^{A}(\omega)~.\label{eqn:eomG<w}
\end{align}
With this, one can derive the following expression for the greater pp Green's function:
\begin{equation}
	\mathcal{G}^>(\omega) = \mathcal{G}^R(\omega)\left[\Sigma^R(\omega) - \Sigma^A(\omega)\right]\mathcal{G}^A(\omega)~.
	\label{<+label+>}
\end{equation}

A direct solution of these equations becomes numerically unstable, as one can already guess from the lesser component of the bare pp Green's function in the diagonal representation,
\begin{equation}
	\mathcal{G}^<_0(\omega)\sim\left(\frac{1}{\omega-\varepsilon+i\eta} -\frac{1}{\omega-\varepsilon-i\eta}\right)e^{-\beta \omega}~,
	\label{eq:G0les}
\end{equation}
where $\varepsilon$ denotes the bare pp energy (atomic eigenenergy). To cancel the divergence of this function for $\omega\rightarrow -\infty$, we introduce a frequency dependent regulator ($\eta\rightarrow \tilde \eta(\omega)$) of the form
\begin{equation}
	\tilde{\eta}(\omega) = \frac{\eta}{1+e^{-\beta\omega}}~,
	\label{eq:regulator}
\end{equation}
following Ref.~\onlinecite{Li2021}. In practice, this damping factor can be incorporated by adding the retarded and lesser self-energies 
\begin{align}
	\Sigma_0^R &= -i\frac{\eta}{2}\left(\tanh\frac{\beta (\omega-\mu_{pp})}{2}+1\right)~,\label{eqn:S0R}\\
	\Sigma_0^< &= i\xi\eta\left(1-\tanh\frac{\beta (\omega-\mu_{pp})}{2}\right)~, \label{eqn:S0<}
\end{align}
where $\mu_{pp}$ is determined such that the pseudoparticle partition function $Z_{pp} =  -i\text{tr}[\xi\mathcal{G}^<(t=0)]$ becomes unity.

\begin{figure}[]
	\centering
	\includegraphics[width=1.0\columnwidth]{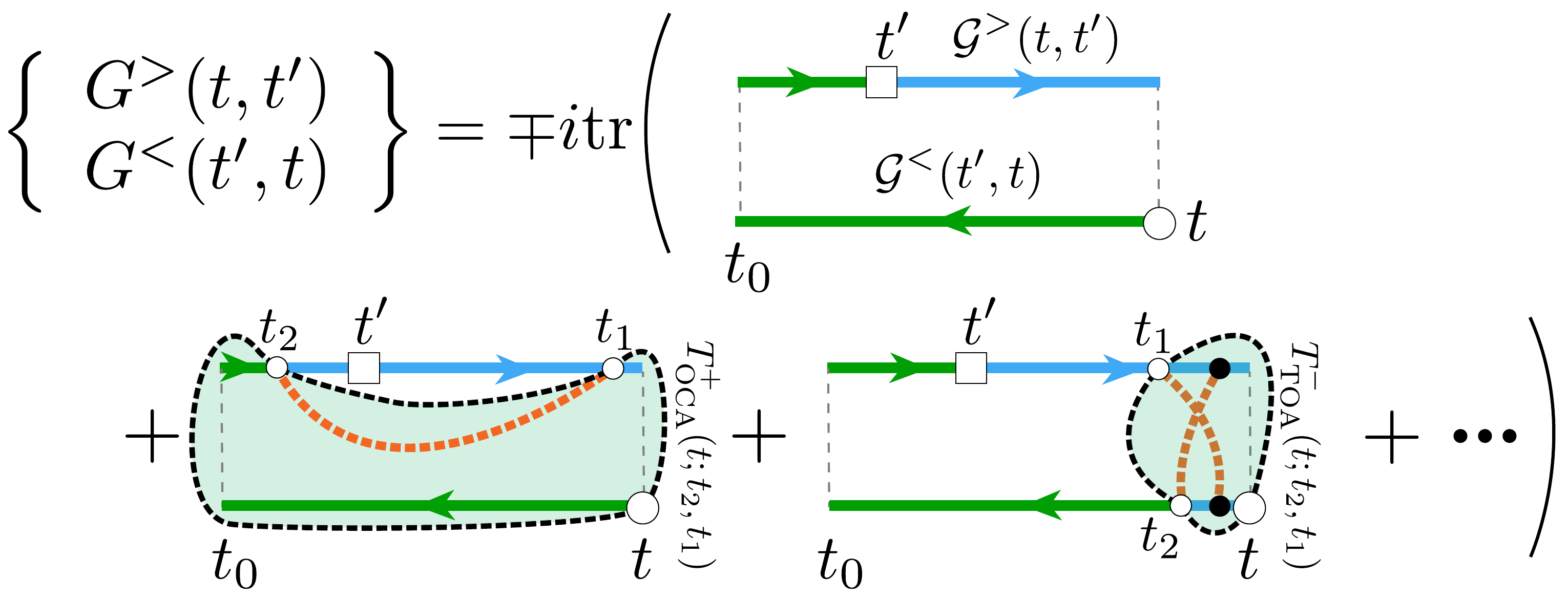}
	\caption{Physical Green's function expressed in terms of the pp Green's functions and hybridization functions (red dashed lines). Blue (green) arrows show greater (lesser) pp Green's functions, and the dashed black line depicts the triangular vertex $T^\pm(F^{(\dagger)}, t; t_2,t_1)$. For $G^>(t,t')$, the annihilation operator is at $t$ and the creation operator at $t'$, while it is the opposite for $G^<(t',t)$.
	}
	\label{fig:phyG}
\end{figure}

Since the pp self-energy depends on the pp Green's functions, the above equations have to be solved self-consistently. Once the converged solution has been found, the physical Green's function $G(t,t')$ of the impurity model can be calculated by evaluating the ring-like diagrams sketched in Fig.~\ref{fig:phyG}. The lesser component $G^<(t',t) = i\langle c^{\dagger}(t)c(t')\rangle$ corresponds to placing the creation operator at $t$ and the annihilation operator at $t'$, while it is the opposite for the greater component $G^>(t,t') = -i\langle c^{}(t)c^{\dagger}(t')\rangle$. 
Like the pp Green's function (or pp self-energy), the physical Green's function depends only on the time difference $t-t'$ in the steady state. 
The first diagram in the figure depicts the NCA contribution  
\begin{align}
	G^>_{\text{NCA}}(t,t')&= \frac{-i}{Z_{pp}}\text{tr}\left[\xi\mathcal{G}^<(t',t)F^{}\mathcal{G}^>(t,t')F^{\dagger}\right]~,\label{Ggtr_nca}\\
	G^<_{\text{NCA}}(t',t)&= \frac{i}{Z_{pp}}\text{tr}\left[\xi\mathcal{G}^<(t',t)F^{\dagger}\mathcal{G}^>(t,t')F^{}\right]~,\label{Gles_nca}
\end{align}
where $\mathcal{G}^{\lessgtr}(t,t')$ is the Fourier transform of the pp Green's function $\mathcal{G}^{\lessgtr}(\omega)$ to the time domain and $F$ is the $4\times 4$ matrix of the operator $c$ acting in the pseudoparticle space. 
The OCA contribution ($n=2$) to the physical Green's function has a vertex correction with a single hybridization line, the $n=3$ contribution a vertex correction with two hybridization lines, etc. Such contributions are sketched in the second row of Fig.~\ref{fig:phyG}.  
To be consistent with the $n$th order self-energy approximation, it is necessary to choose $n-1$ hybridization lines for the vertex correction.
We can combine these vertex corrections into a triangular vertex function $T^\pm(F^{(\dagger)},t; t_2,t_1)$ and express the Green's function contributions beyond NCA as ($t_0<t'<t$)
\begin{align}
G^>_\text{vert.}(t,t')=& -\frac{i}{Z_{pp}}\int_{t'}^t dt_1 \int_{t_0}^t  dt_2~\text{tr}\Big[\xi\mathcal{G}^<(t',t_2)\nonumber\\
&\hspace{8mm}\times T^-(F^{},t; t_2,t_1)\mathcal{G}^>(t_1,t')F^{\dagger}\Big]\nonumber\\
&+\frac{i}{Z_{pp}}\int_{t'}^t dt_1 \int_{t_0}^{t'}  dt_2~\text{tr}\Big[\xi\mathcal{G}^>(t',t_2)\nonumber\\
&\hspace{8mm}\times T^+(F^{},t; t_2,t_1)\mathcal{G}^>(t_1,t')F^{\dagger}\Big]
\label{Ggtr_oca}, 
\end{align}
and similarly for the lesser component. Here, $t_0$ denotes the left end of the considered time interval, and the superscript of the $T$ vertex indicates whether the $F^{(\dagger)}$ operator at $t_2$ is located on the backward  ($-$) or forward ($+$) branch.

\subsection{Calculation of \texorpdfstring{$\Sigma$}{Sigma} with QTCI}
\label{sec_qtci}

In the calculation of the pseudoparticle self-energies and physical Green's functions for $n>1$, the internal time variables are integrated over, respecting their cyclic ordering along the contour \cite{Eckstein2010}. We now explain how to calculate these integrals efficiently using QTCI, focusing on the OCA self-energy diagrams. In this case, we can skip the definition of a $T$ vertex and directly fit the pp self-energy. 

\begin{figure}[]
	\centering
	\includegraphics[width=1.0\columnwidth]{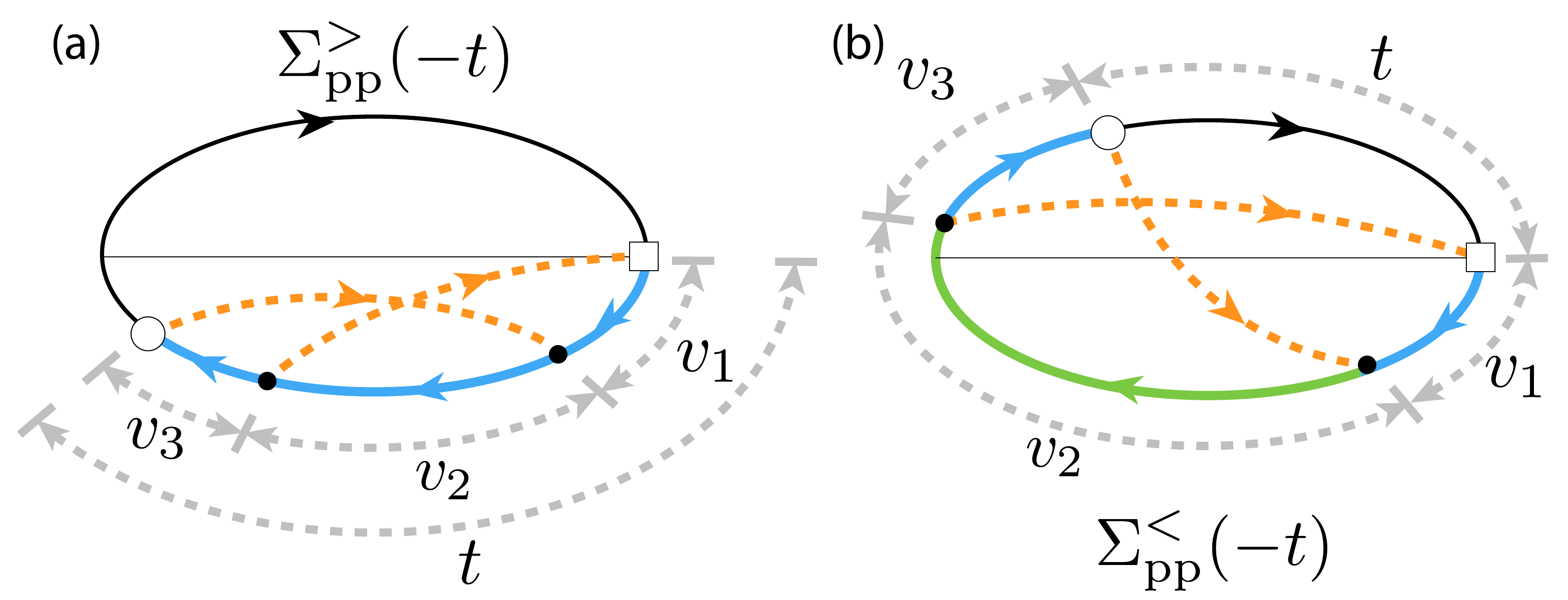}
	\caption{OCA diagrams for the pp self-energy on the ring-shaped real-time contour. Here the left-most point represents physical time $t=0$ and the rightmost point $t=t_\text{max}$. Panel (a) corresponds to $\Sigma^>(-t)$ and panel (b) to $\Sigma^<(-t)$. The gray dashed lines indicate the time difference variables $v_1$, $v_2$, $v_3$ and the time argument $t$. Greater (lesser) pp Green's functions are represented by the blue (green) arrows. Only one of the four OCA diagrams is shown. The orientations of the hybridization lines in panel (b) correspond to the expression given in the text. 
	}
	\label{Fig:ringOCA}
\end{figure}

We consider the forward and backward branches of the Keldysh contour in the real-time interval from $t=0$ to $t_\text{max}$, and fuse these branches into a closed contour (Fig.~\ref{Fig:ringOCA}). 
Along the closed ring-shaped contour, one can conveniently represent both the greater and the lesser components of the self-energy with the same initial time point;
the first operator of the pseudoparticle self-energy is placed at $t_\text{max}$ and the last one on the lower branch for the greater component [Fig.~\ref{Fig:ringOCA}(a)] or upper branch for the lesser component [Fig.~\ref{Fig:ringOCA}(b)]. Because of the cyclic order of the times on this fused contour, it is convenient to define time-difference variables for a contour time $0\le v\le 2t_\text{max}$, which runs from $t=t_\text{max}$ ($v=0$) along the lower branch to $t=0$ ($v=t_\text{max}$) and then along the upper branch back to $t=t_\text{max}$ ($v=2t_\text{max}$). Defining the contour time differences between the vertices by $v_i$ ($i=1,2,3$), the OCA self-energy contribution can be written as
\begin{align}
	\Sigma^a_{\mathrm{OCA}}(-t) =& \int_{0}^{2t_{\mathrm{max}}}dv_1\int_{0}^{2t_{\mathrm{max}}}dv_2\int_{0}^{2t_{\mathrm{max}}}dv_3\nonumber\\
	&\hspace{0.cm} \times \tilde{\delta}\left(a;t,\sum^{}_{i}v_i\right)\sigma(v_1,v_2,v_3)~,
	\label{eqn:QTCI_OCA}
\end{align}
where $a$ labels the greater and lesser component, the $\tilde{\delta}$ function imposes a constraint on the sum of the difference variables, 
\begin{align}
	\tilde{\delta}\left(>;t,\sum^{}_{i}v_i\right) &= \delta\left(t-\sum^{}_{i}v_i\right)~,\\
	\tilde{\delta}\left(<;t,\sum^{}_{i}v_i\right) &= \delta\left(2t_{\mathrm{max}}-t-\sum^{}_{i}v_i\right)~,
	\label{<+label+>}
\end{align}
and $\sigma(v_1,v_2,v_3)$ denotes the matrix-valued weight of the OCA diagram with the corresponding operator positions. Specifically, for the diagram shown in Fig.~\ref{Fig:ringOCA}(b) this weight is 
\begin{align}
&\sigma(v_1,v_2,v_3)=\sum_{\mu\nu} F_\nu\mathcal{G}^>(v_3)F_\mu\mathcal{G}^<(-2t_\text{max}+2v_1+v_2)\nonumber\\
&\hspace{5mm}\times F_\nu^\dagger\mathcal{G}^>(-v_1)F_\mu^\dagger\nonumber\\
&\hspace{5mm}\times\Delta^{>}_\mu(2t_\text{max}-v_1-v_2)
\Delta^>_\nu(2t_\text{max}-2v_1-v_2-v_3)\nonumber\\
&\hspace{5mm}+\Big(\ldots\Big)~,
\label{eq_integrand}
\end{align}
where the sum is over the spin flavors of the hybridization functions $\Delta$ and the parenthesis stands for the three additional terms with different directions of the hybridization lines. 

\begin{figure}[]
	\centering
	\includegraphics[width=1.0\columnwidth]{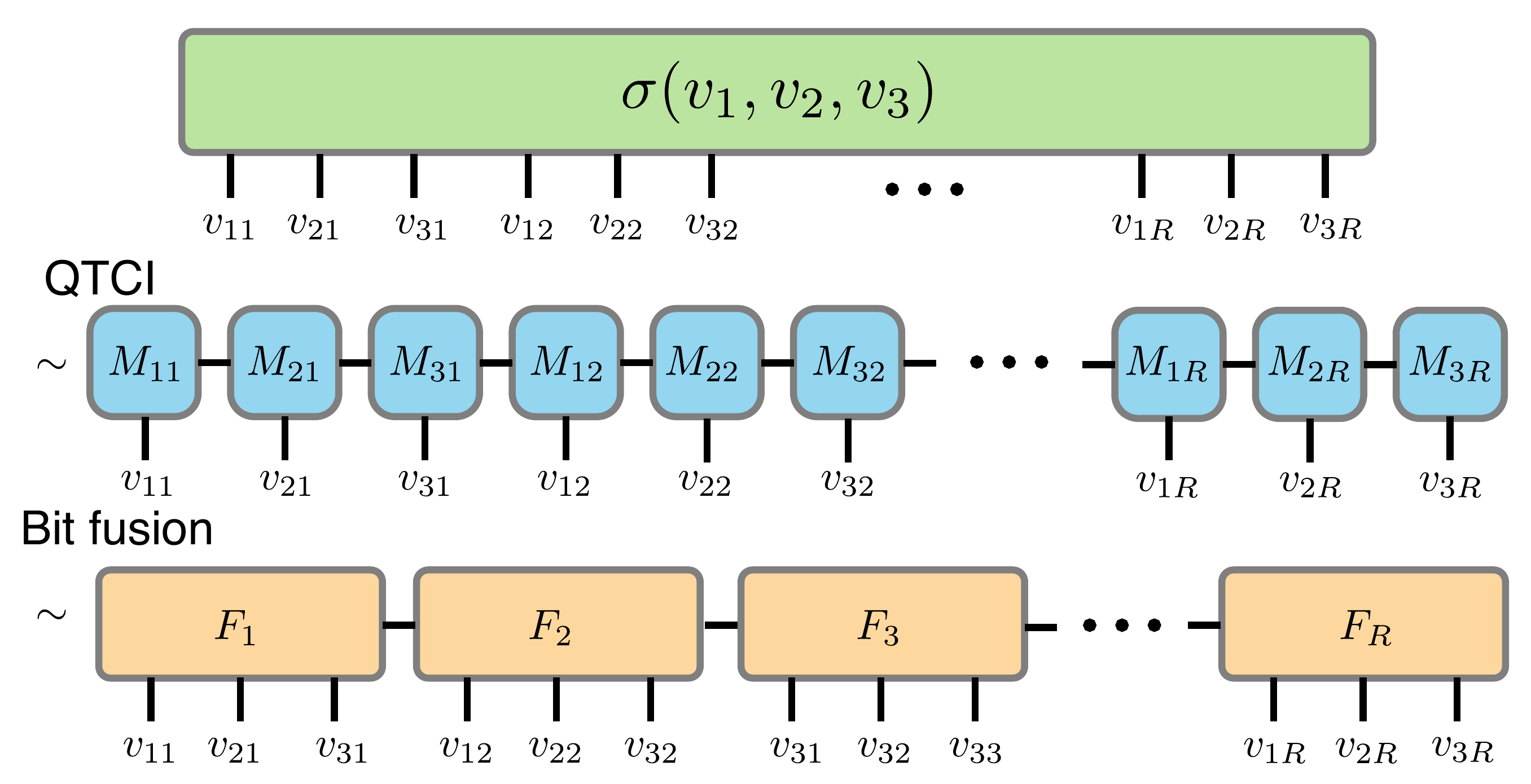}
	\caption{Tensor-train representation of $\sigma(v_1,v_2,v_3)$. The top panel illustrates the function in the quantics representation. The middle panel shows the tensor train obtained by the QTCI factorization, and the bottom panel the tensor train with fused legs. 
	}
	\label{Fig:tensorTrain}
\end{figure}

Next, the variables $v_i$ are expressed in a binary representation with $R$ digits \cite{Shinaoka2023}, 
\begin{equation}
v_i = \sum^{R}_{j=1} v_{ij}\frac{2t_{\text{max}}}{2^j},
\end{equation}
which yields the quantics representation of the function $\sigma(v_1,v_2,v_3)$. This function is then decomposed into a tensor form by the QTCI method~\cite{Ritter2024} (Fig.~\ref{Fig:tensorTrain}).
We use the open source code xfac \cite{Fernandez2024} in our implementation. 
The quality of the QTCI fit is generally excellent for the greater component. The lesser component is more challenging, since every time an internal integral shifts an operator from the lower to the upper branch at the left end of the ring-shaped contour, the integrand exhibits a discontinuity. This could be fixed by splitting the lesser self-energy into integral contributions with one, two and three operators on the upper branch and evaluating those separately. However, QTCI should in principle be capable of fitting functions with discontinuities. We thus explore here a different strategy and fuse bits of the same scale in the binary representation (bottom panel of Fig.~\ref{Fig:tensorTrain}). We show in the results section that the fused bit tensor representation can fit the underlying function with high accuracy. It appears that the fused bits of the largest scale approximately classify the continuous domains of the function, eliminating the need for a reparametrization of the integral with explicit sum over branch indices. Whether or not this is possible also at higher orders (for $n>2$) remains to be tested.

From Eq.~(\ref{eqn:QTCI_OCA}) we can derive a formula for the direct evaluation of the pp self-energy in the frequency domain.
For example, rewriting the $\delta$ function in the expression for the greater component in the Fourier representation,
\begin{equation}
	\delta\Bigg(t-\sum^{}_{i}v_i\Bigg) = \frac{1}{2\pi}\int_{-\infty}^{\infty}d\omega~e^{i\omega(t-\sum^{}_{i}v_i)}~,
	\label{eqn:delta_function}
\end{equation}
we obtain
\begin{align}
	\Sigma^>_{\mathrm{OCA}}(-t) =& \frac{1}{2\pi}\int_{-\infty}^{\infty}d\omega~e^{+i\omega t}\int_{0}^{2t_{\mathrm{max}}}dv_1\int_{0}^{2t_{\mathrm{max}}}dv_2 \nonumber\\
	& \times \int_{0}^{2t_{\mathrm{max}}}dv_3 e^{-i\omega\sum^{}_{i}v_i}\sigma(v_1,v_2,v_3)~.
	\label{<+label+>}
\end{align}
Comparing this to the Fourier transform of the pp self-energy, $\Sigma^>_{\mathrm{OCA}}(-t) = -\frac{1}{2\pi}\int_{-\infty}^{\infty}d\omega~e^{+i\omega t}\Sigma^A_{\mathrm{OCA}}(\omega)$, one finds the frequency expression
\begin{align}
\Sigma^>_{\mathrm{OCA}}(\omega) &= -2i\mathrm{Im}\Sigma^A_\text{OCA}(\omega)\nonumber\\
	&= 2i\mathrm{Im}\bigg[\int_{0}^{2t_{\mathrm{max}}}dv_1\int_{0}^{2t_{\mathrm{max}}}dv_2\int_{0}^{2t_{\mathrm{max}}}dv_3\nonumber\\
	&\hspace{15mm}\times e^{-i\omega\sum^{}_{i}v_i}\sigma(v_1,v_2,v_3)\bigg]~.
	\label{<+label+>}
\end{align}
Since the self-energy is only measured for negative times on the ring-shaped contour, we calculate the greater and lesser Green’s functions via the advanced Green’s function in frequency space.
In the quantics tensor-train representation, the function $\sigma(v_1,v_2,v_3)$ becomes a product of tensors. In the nonfused representation, we have $\sigma^{\text{QTCI}}(\left\{v_{ij}\right\})=M_{11}(v_{11})\cdot M_{21}(v_{21}) \cdot M_{31}(v_{31}) \cdot M_{12}(v_{12}) \cdots M_{3R}(v_{3R})$ and in the fused representation 
$\sigma^{\text{QTCI}}_\text{fused}(\left\{v_{ij}\right\})=F_1(v_{11},v_{21},v_{31})\cdot F_2(v_{12},v_{22},v_{32}) \cdots F_R(v_{R1},v_{R2},v_{R3})$, see Fig.~\ref{Fig:tensorTrain}. Merging the integration measures into a prefactor $\mathcal{N}$, we can therefore write the result in the nonfused representation and for general order $n$ as 
\begin{align}
	\Sigma^>(\omega) = 2i\mathrm{Im} \Bigg[&\mathcal{N}\left(\sum^{1}_{v_{11}=0}W(v_{11};\omega)M_{11}(v_{11})\right)\nonumber\\
	&\hspace{0mm}\times\left(\sum^{1}_{v_{21}=0}W(v_{21};\omega)M_{21}(v_{21})\right)
         \times \dots
	\label{<+label+>}\Bigg]~, 
\end{align}
where the weight factors $W$ are given by 
\begin{equation}
W(v_{ij}) = \exp\left\{-i\omega 2t_{\text{max}}\frac{v_{ij}}{2^{j}}\right\}~.
\end{equation}
A similar expression (with more terms in each factor) is obtained for the fused representation. 
The computational cost of this measurement is $\mathcal{O}(N_\omega R (2n-1) N_\text{bond}^3)$, where $N_\omega$ is the number of measured frequency points, $R$ is the number of digits in the binary encoding, $n$ ($=2$ for OCA) is the order of the expansion, and $N_\text{bond}$ is the maximum bond dimension of the tensor train. 

With the above procedure, we directly compute the frequency-dependent self-energy needed in the self-consistent calculation described in Sec.~\ref{sec_pp}. 

\subsection{Calculation of the physical Green's functions with QTCI}
\label{subsec_GF}

The greater component of the physical Green's function is obtained by summing the contributions \eqref{Ggtr_nca} and \eqref{Ggtr_oca}, and a similar calculation yields the lesser component. 
The NCA contribution can be evaluated directly, while the vertex correction involves two integrals which can be computed with QTCI. We briefly describe the procedure for the greater component [Eq.~\eqref{Ggtr_oca}].  

As in the case of the pp self-energy computation, the physical Green's function can be written as an integral over three variables $v_i$ ($i=1,2,3$) on the ring contour,  
which we define in a clock-wise fashion starting from $t_\text{max}$, analogous to Fig.~\ref{Fig:ringOCA}. Specifically, $v_1$ encodes the location of $t_2$ (on the lower branch for $0<v_1<t_\text{max}$ and on the upper branch for $v_1>t_\text{max}$), $v_1+v_2$ the location of $t'$ and $v_1+v_2+v_3$ the location of $t_1$ in Fig.~\ref{fig:phyG}. With this parametrization, we can write the greater Green's function for $t=t_\text{max}-t'$ as 
\begin{align}
	&G_{\text{vert}}^{>}(t) = \int_{0}^{2t_{\text{max}}}dv_1 \int_{0}^{2t_{\text{max}}}dv_2 \int_{0}^{2t_{\text{max}}}dv_3~\nonumber\\
	&\hspace{5mm}\times  \delta(2t_\text{max}-v_1-v_2-t)g^{>}_\text{vert}(v_1,v_2,v_3)~,\label{eq_gvert}\\
	&g^{>}_\text{vert}(v_1,v_2,v_3) = -\frac{i}{Z_{pp}}\text{tr}\Big[\xi\mathcal{G}(v_1+v_2,v_1)\nonumber\\
	&\hspace{5mm}\times T(F^{},2t_\text{max}-v_1-v_2;v_1,v_1+v_2+v_3) 
	\mathcal{G}^>(v_3)
	F^{\dagger}\Big]~, 
	\label{ggtr_oca}
\end{align}
where in the trace the first two factors are $\mathcal{G}^>T^+$ for $v_1>t_\text{max}$ or $\mathcal{G}^<T^-$ for $v_1<t_\text{max}$.
Here, the assumption is that the pp Green's functions and hybridization functions are sufficiently decayed for relative times $\gtrsim~t_\text{max}$.
We choose $t>0$, since the Green's function for negative times can be obtained from the relation $G^{>,<}(t) = -[G^{>,<}(-t)]^*$. 
Using the Fourier representation of the $\delta$ function as in Eq.~(\ref{eqn:delta_function}), but this time only with the sum over $v_1$ and $v_2$, one can directly measure the physical Green's function in $\omega$ space.

\begin{figure}[t] 
\begin{centering}
\includegraphics[width=0.9\columnwidth]{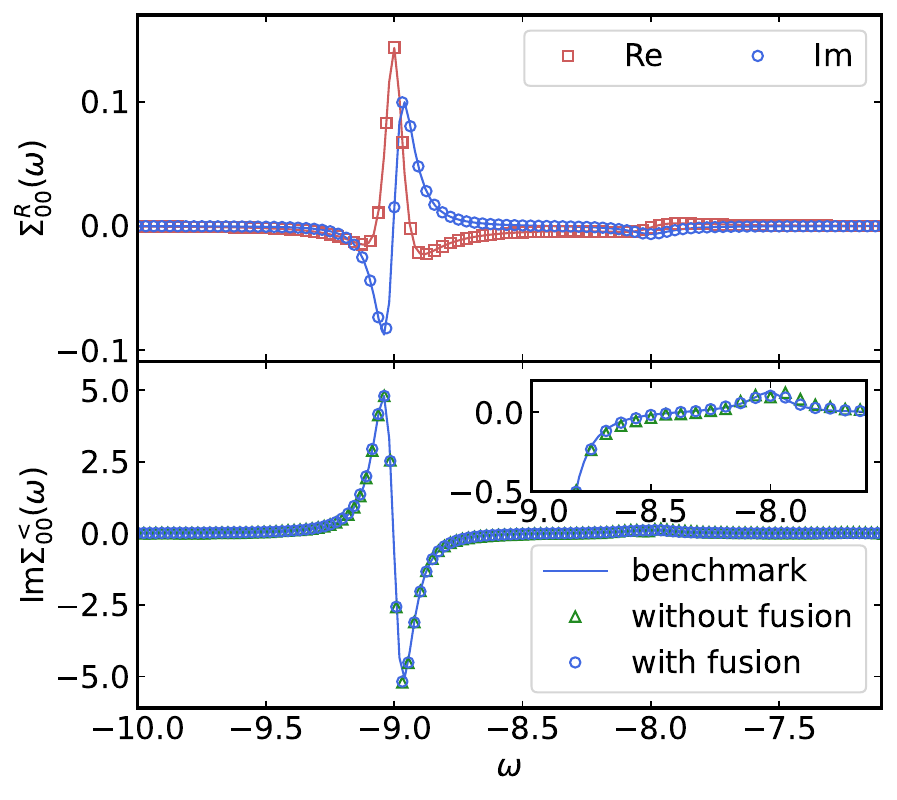}
\par\end{centering}
\caption{
Retarded and lesser components of the one-shot OCA pp self-energy for the single bath model with $U = 10$, $\mu = 5$, $g_\text{bath} = 1$, $\epsilon_\text{bath} = 4.0$, and $T = 1$ (empty state 00, $\eta=0.1$). The line shows the exact reference data, while the squares and circles show the QTCI fits obtained in the fused-leg representation ($F$ tensors in Fig.~\ref{Fig:tensorTrain}). In the bottom panel, the triangles show the less accurate QTCI fit in the original quantics representation ($M$ tensors in Fig.~\ref{Fig:tensorTrain}).}
\label{Fig:testfused}
\end{figure}

In the case of OCA, the vertex $T^\pm$ is the sum of two terms, each of which involves a product of a hybridization function, two pp Green's functions and creation and annihilation operators  (Fig.~\ref{fig:phyG}). It is not necessary in this case to precompute and store $T$, since it can be calculated on the fly. 

\begin{figure*}[t] 
\begin{centering}
\includegraphics[width=\textwidth]{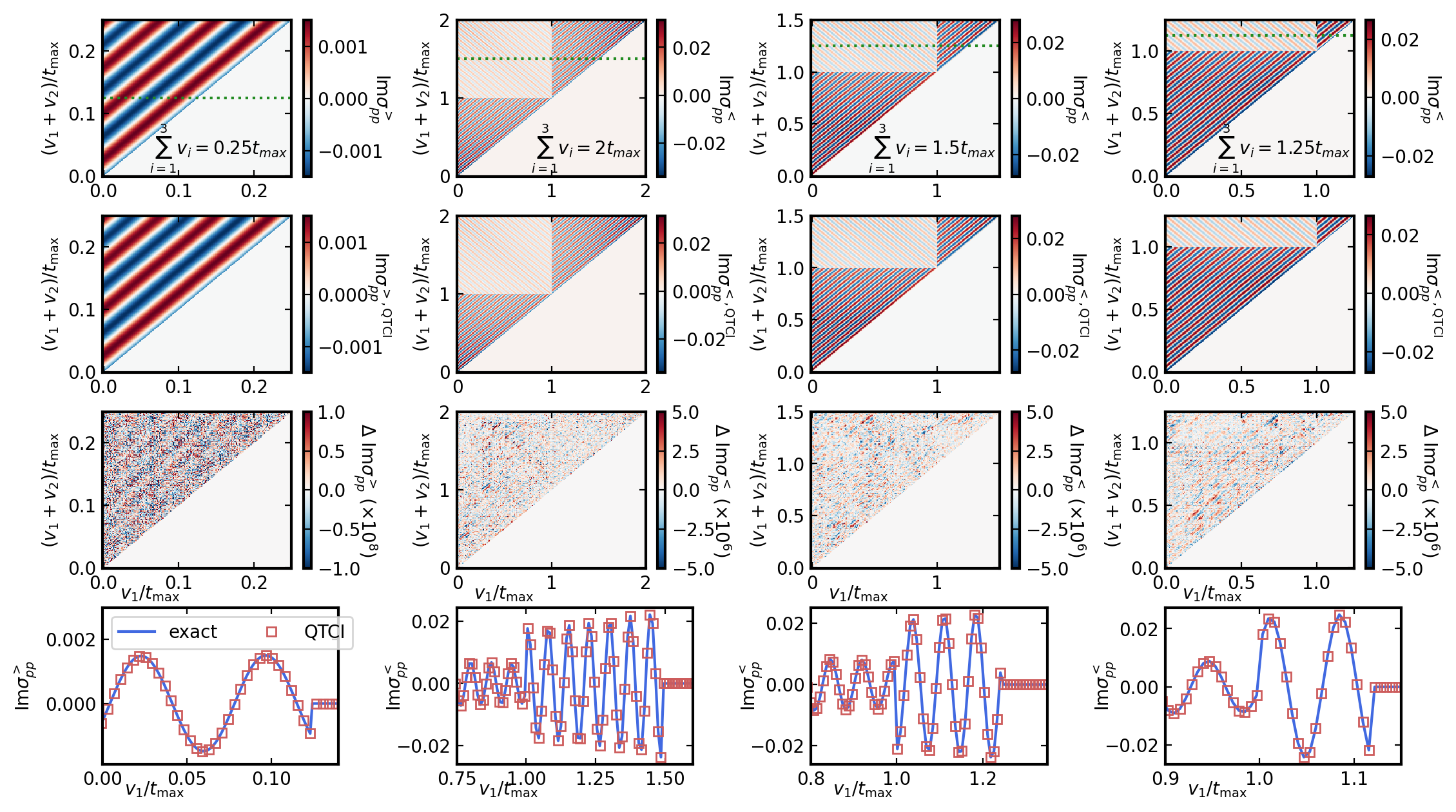}
\par\end{centering}
\caption{
Greater and lesser components of the integrand of the one-shot OCA pp self-energy for the single bath model with $U = 10$, $\mu = 5$, $g_\text{bath} = 1$, $\epsilon_\text{bath} = 4.0$, $T = 1$ and small damping $\eta=0.01$. The left column is for $v_1+v_2+v_3=0.25t_\text{max}$ ($t_{\text{max}} = 84.7$), which corresponds to the greater component. The remaining columns show slices of the lesser component with $v_1+v_2+v_3=2t_\text{max}$, $1.5t_\text{max}$ and $1.25t_\text{max}$. The top row shows the reference data, the second row the (fused-leg) QTCI fits, and the third row the difference between the two.
The bottom row presents a comparison between the exact and the QTCI results along the cuts with the fixed $v_1+v_2$ indicated by the green dotted lines in the top panels. 
Only a subset of points is shown for a better visualization.
}
\label{Fig:testqtci}
\end{figure*}

\section{Results}
\label{sec:results}

\subsection{QTCI of discontinuous integrand for the pp self-energy: AIM with single bath site}

We first show some test calculations for an Anderson impurity model with a single bath site. Without the damping factors in Eqs.~\eqref{eqn:S0R} and \eqref{eqn:S0<}, this is a challenging case for the hybridization expansion, because the hybridization function does not decay in time. With a strong damping factor, the situation is similar to a model with continuous bath spectrum. The physical parameters of the model are: impurity interaction $U = 10$, chemical potential $\mu = 5$, bath coupling $g_\text{bath} = 1$, bath energy $\epsilon_\text{bath} = 4.0$, and temperature $T = 1$. 
Figure~\ref{Fig:testfused} presents the retarded and lesser components of the pp self-energy for a one-shot OCA calculation (empty state 00, OCA contribution only) with damping $\eta=0.1$ in Eq.~\eqref{eq:regulator}.
Here, the pp self-energy diagram is composed of noninteracting pp Green's functions $\mathcal{G}_0$.
The lines show the reference data from an exact numerical evaluation of the integrals and the circles the QTCI data. A relatively coarse frequency grid is used in order not to clog the figure. In the panel for the lesser component, we compare the original and fused-leg versions of the QTCI fit, with the inset showing a zoom of the frequency region near $\omega=-8$. The fused-leg version ($F$ tensors) captures the small dip in the self-energy more accurately than the original QTCI fit ($M$ tensors). We will thus focus in the remaining subsections on data obtained with the fused-leg representation.  

The top row of Fig.~\ref{Fig:testqtci} shows the imaginary part of the integrand \eqref{eq_integrand} as a function of $v_1$ and $v_2$, for the indicated fixed values of $v_1+v_2+v_3$. These are the reference data $\sigma_\text{reference}$ that we aim to factorize with the QTCI fit. The first column, with $v_1+v_2+v_3=0.25t_\text{max}$, corresponds to a greater contribution ($\sigma^>$) and the remaining columns, with $v_1+v_2+v_3=2t_\text{max}$, $1.5t_\text{max}$ and $1.25t_\text{max}$ to a lesser one ($\sigma^<$). The second row of the figure shows the (fused-leg) QTCI fit of the function, $\sigma_\text{QTCI}$, and the third row the difference $\Delta\sigma=\sigma_\text{reference}-\sigma_\text{QTCI}$. Here, in order to obtain a good fit, we have placed six initial pivots near the discontinuities of the function ($v_1=t_\text{max}$ and $v_1+v_2=t_\text{max}$) within the bulk of the three-dimensional integration domain. With this initial condition, the QTCI fit achieves a good accuracy, both for the greater component (without discontinuities) and the lesser component (with discontinuities). The largest absolute values of $\Delta\sigma$ are four orders of magnitude smaller than the largest absolute values of $\sigma$. Furthermore, $\Delta\sigma$ looks almost like random noise, whose contribution to the integrated quantity $\Sigma$ will be very small. 

To illustrate the QTCI's ability to resolve the discontinuities in $\sigma(v_1,v_2,v_3)$, we show in the bottom row of Fig.~\ref{Fig:testqtci} some cuts marked by the green dashed lines in the top row. For a better visualization, only a subset of the $2^{18}=262144$ data points is shown.

\subsection{Equilibrium pp Green's functions of the AIM with continuous bath}

\begin{figure}[t]
	\centering
	\includegraphics[width=0.9\columnwidth]{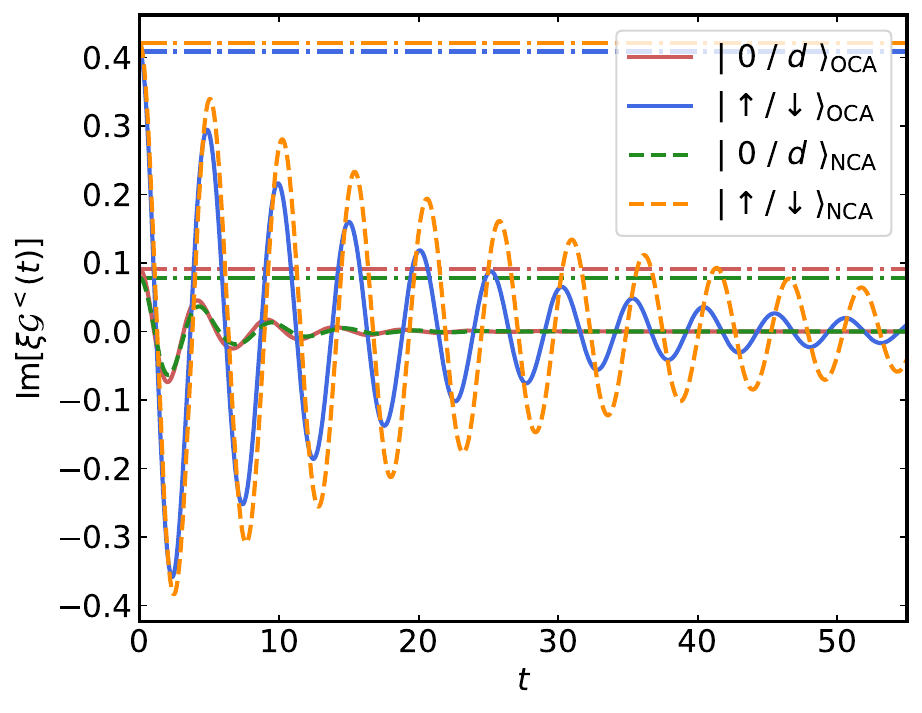}
	\caption{
		Pseudoparticle Green's functions of the Anderson impurity model with semicircular bath.
		The parameters are $T=0.1, U=2$, $\mu=1$ and the bath coupling strength is $g=0.5$.
		The damping factor is $\eta=0.01$. Full (dashed) lines show the OCA (NCA) results. 
		The horizontal dash-dotted lines indicate $-\mathcal{G}_\alpha^M(-i\beta)$ obtained from an independent Matsubara-axis code.
	}
	\label{Fig:test_pp_oca}
\end{figure}

We next show in Fig.~\ref{Fig:test_pp_oca} the interacting pp Green's functions in the time domain, $\text{Im}\xi^{}_\alpha\mathcal{G}_\alpha^<(t)$, with $\alpha=\{0,\uparrow,\downarrow,\uparrow\downarrow\,\,\equiv d\}$, for an impurity model with a semicircular bath density of states $\rho_\text{bath}$.  
Here, we use the half-bandwidth of the bath as the unit of energy (or inverse time).
The model parameters are: $T=0.1, U=2$, $\mu=1$ and the bath coupling strength is $g=0.5$ [$\Delta^R(\omega)=g^2\int d\epsilon \rho_\text{bath}(\epsilon)/(\omega-\epsilon+i0^+)$]. The dashed lines show the NCA results and the solid lines the sum of the NCA and OCA contributions. Since the system is particle-hole symmetric and there is no magnetic field, the results for the doubly occupied and empty pp Green's functions, as well as for the $\alpha=\,\uparrow,\downarrow$ pp Green's functions are identical. We further notice that the pp Green's functions for the empty and doubly occupied states decay fast, whereas those for the singly occupied states decay slowly (here controlled by the relatively large $\eta=0.01$). It is thus important to use a large enough $t_\text{max}$.

The horizontal dash-dotted lines indicate the Matsubara pp Green's function values $-\mathcal{G}_\alpha^M(-i\beta)$, obtained from an independent Matsubara-axis code \cite{Kim2022,noVertex}~. These values should be identical to $\text{Im}\xi^{}_\alpha\mathcal{G}_\alpha^<(t=0)$. The results match well up to small discrepancies which can be explained by the regularization parameter $\eta$ in the steady-state formulation. 
The OCA correction leads to a slight enhancement of the double occupation ($\frac{-i}{Z_{pp}}[\xi\mathcal{G}^<(t=0)]_{\alpha=d}$), compared to the NCA result. This is consistent with the overestimation of the correlation effects in NCA \cite{Eckstein2010}.  

\begin{figure}[t]
	\centering
	\includegraphics[width=0.9\columnwidth]{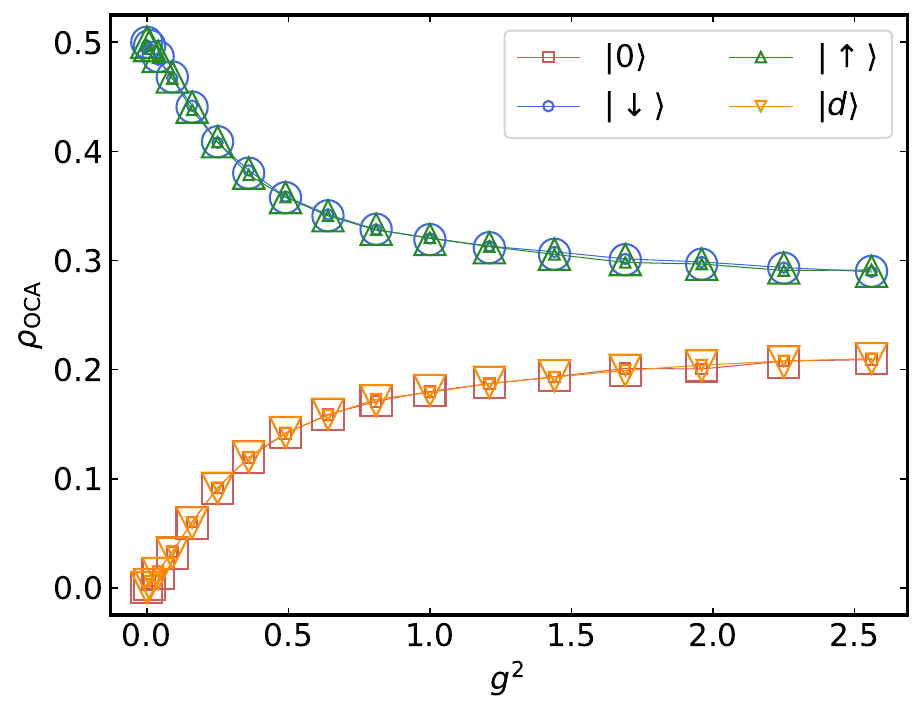}
	\caption{
		Local density matrix of the Anderson impurity model with semicircular bath density of states.
		The figure shows the OCA result for $T=0.1, U=2, \mu=1$.
		The damping factor is $\eta=0.01$.
		Large symbols show the results from the Matsubara-axis code, and small symbols the results from the QTCI-based steady-state code. 
	}
	\label{Fig:test_densitymatrix}
\end{figure}

Figure~\ref{Fig:test_densitymatrix} plots the probabilities of the different eigenstates of $H_\text{loc}$, or diagonal elements of the density matrix $\rho_{\alpha}=\frac{-i}{Z_{pp}}[\xi\mathcal{G}^<(t=0)]_\alpha$, as a function of the coupling strength $g$ to the bath. The large square, circle, and (inverted) triangle symbols are the results from the Matsubara-axis code, while the small symbols indicate the results from the QTCI-based steady-state calculation with $\eta=0.01$. In the limit of large bath coupling strength, the hybridization term dominates and the correlation effects are suppressed. 
This explains why the probabilities for doublons (d), holons (0), and singlons ($\uparrow,\downarrow$) approach each other with increasing $g$.

\subsection{QTCI of equilibrium physical Green's functions: AIM with continuous bath}

\begin{figure}[t]
	\centering
	\includegraphics[width=1.0\columnwidth]{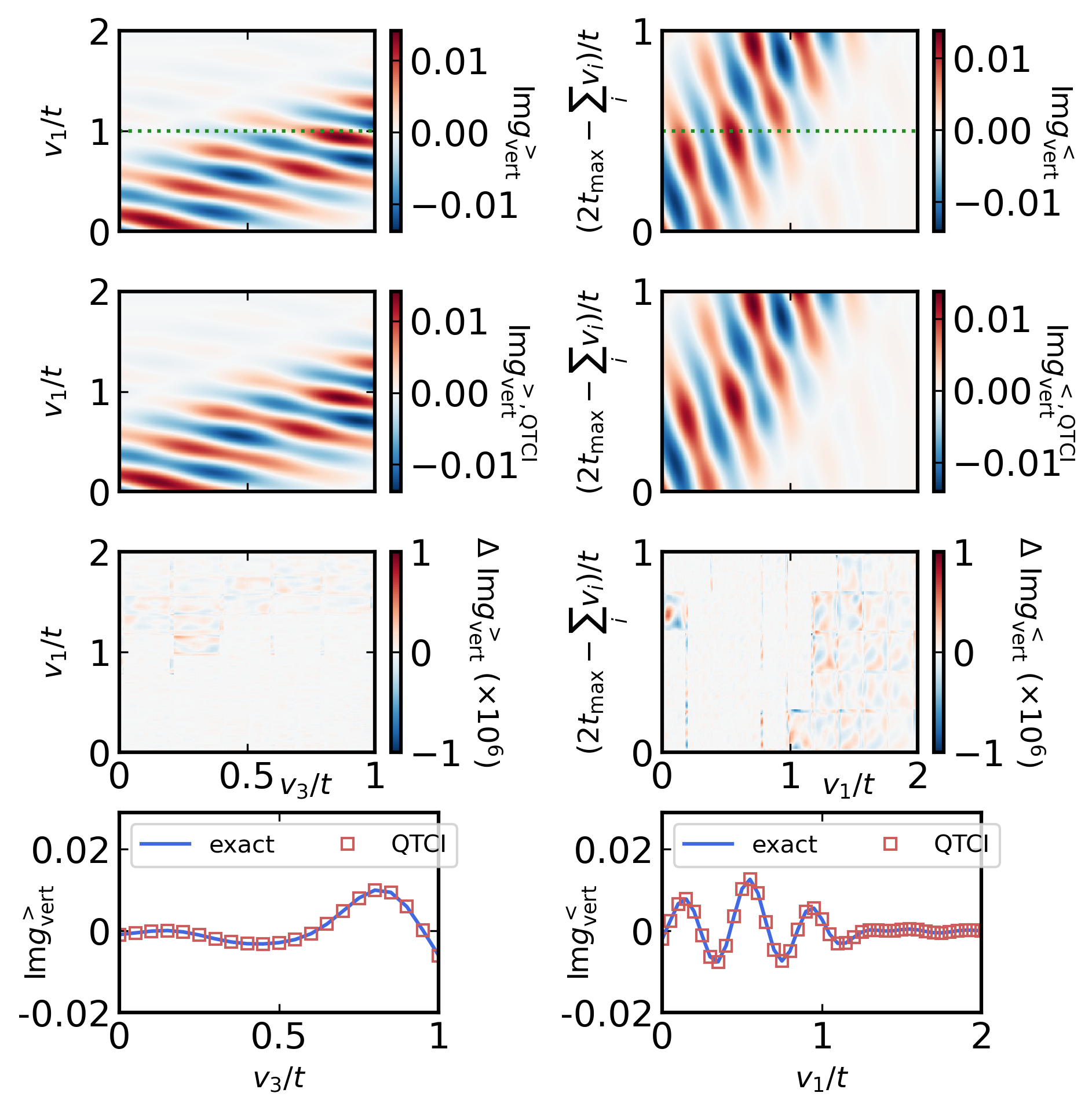}
	\caption{
		QTCI approximation for the integrand of the physical Green's function, $g_{\text{vert}}(v_1,v_2,v_3)$.
		The pp Green's functions constituting $g_{\text{vert}}$ are the self-consistent OCA solutions of the Anderson impurity model with semicircular bath for $T=0.1, U=2, \mu=1.0, g_\text{bath}=0.5, \eta=0.01$~.
		The figure shows a slice for $t\sim 8.04$. 
		The original data for the imaginary parts of the greater and lesser components are plotted in the first row, the QTCI fits in the second row, and the fitting error is shown in the third row. 
		Cuts along the green dotted lines in the top panels (with only a subset of points for better visualization) are shown in the bottom panels. 
	}
	\label{fig:phyG_qtci_fit}
\end{figure}

We calculate the physical Green's function from the pp Green's functions and hybridization functions using a QTCI factorization of the integrand (\ref{ggtr_oca}). 
The quality of the QTCI fit is illustrated in Fig.~\ref{fig:phyG_qtci_fit} for the Anderson impurity model with semicircular bath density of states, whose half bandwidth is used as the unit of energy. 
The parameters are $T=0.1$, $U=2$, $\mu=1.0$, $g_\text{bath}=0.5$,  and $\eta=0.01$. 
The figure compares the greater and lesser components of $\text{Im}g_\text{vert}$ as a function of the variables $v_1$ and $\sum^{}_{i}v_i$ for fixed relative time $t=8.04$. 
The top row of the figure shows the original function, the second row the fitted function, and the third row the difference between the two. 

Also $g_\text{vert}$ can be fitted with four digits accuracy and the fitting error is essentially random noise, so that the integrated quantity (physical Green's function) is well reproduced by the QTCI approach. In the bottom panels, we show the direct comparison of the original and fitted function along the cuts indicated by the green dotted lines in the top row. Only a subset of the $2^{18}=262144$ data points is shown for a better visualization.

\subsection{DMFT solutions for the Hubbard model on the infinite-dimensional Bethe lattice}

In this section, we use DMFT and the QTCI-based strong-coupling solver to calculate equilibrium and nonequilibrium spectral functions of the Hubbard model on an infinitely connected Bethe lattice. For this system, the DMFT self-consistency condition becomes \cite{Georges1996}
\begin{equation}
\Delta(t,t')=(\tfrac12 D)^2 G(t,t'), \label{eq_selfconsistency}
\end{equation}
where $D$ is the half-bandwidth of the noninteracting density of states, which we use as the unit of energy. 

\begin{figure}[t]
	\centering
	\includegraphics[width=1.0\columnwidth]{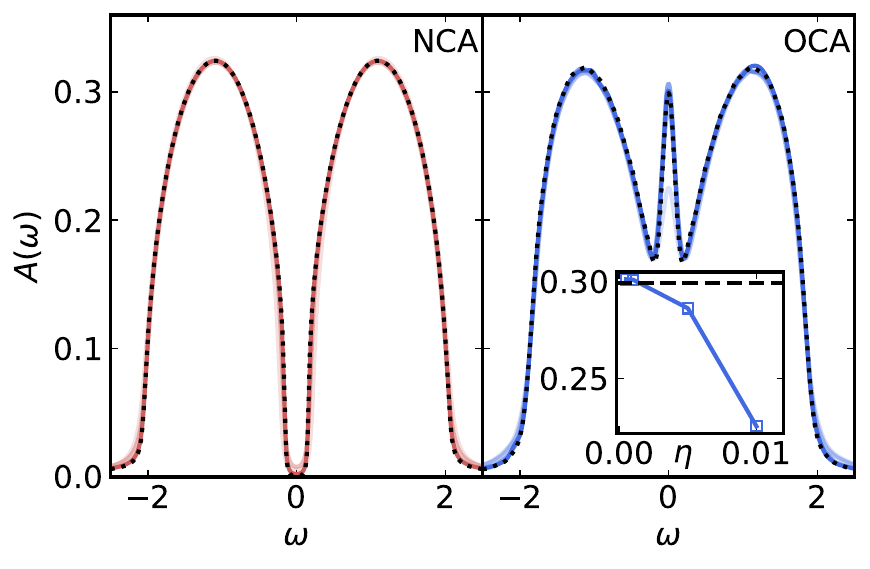}
	\caption{
		Equilibrium spectral function of the half-filled Hubbard model on the infinite-dimensional Bethe lattice for $T=0.1$, $U=2$.
		The left (right) panel shows the NCA (OCA) results with various damping factors: from faint to vivid color, $\eta=0.01, 0.005, 0.001, 0.0005$. 
		The inset of the OCA panel plots $A(\omega=0)$ as a function of $\eta$ (symbols) and compares it to the reference value (horizontal dashed line).
		The black dashed lines show the reference spectra obtained from a calculation on the L-shaped Kadanoff-Baym contour.
	}
	\label{fig:dmft_benchmark}
\end{figure}

\subsubsection{Equilibrium system}

Figure~\ref{fig:dmft_benchmark} plots the NCA and OCA equilibrium spectral functions $A(\omega)=-\frac{1}{\pi}\text{Im}G^R(\omega)$ for $U=2$, $T=0.1$ and different damping parameters $\eta$. For small $\eta$, the solutions from the real-time/frequency solver approach the reference data obtained with a conventional implementation of the NCA and OCA solutions on the L-shaped Kadanoff-Baym contour \cite{Eckstein2010}.  

$U=2$ is on the metallic side of the Mott transition line of this model and $T=0.1$ above the critical end-point, so the system is in the metal-insulator crossover regime which features a (suppressed) quasi-particle peak \cite{Bluemer2005}. The NCA solution instead exhibits a gap, since the Mott transition line in this approximation is shifted to substantially lower $U$ \cite{Eckstein2010}. The OCA solution yields a metallic spectrum with a small quasi-particle peak, qualitatively consistent with the expected result. 

For $\eta=0.01$, the damping effect leads to inaccurate spectra, and in particular suppresses the quasi-particle peak in the OCA solution. 
Both the NCA and OCA results however approach the reference data quantitatively as the $\eta$ value is decreased. The insulating NCA solution converges fast with decreasing $\eta$, while the metallic OCA solution converges more slowly. In the inset of the right panel, we plot the height of the quasi-particle peak, $A(\omega=0)$, as a function of $\eta$. The symbols are the results of the QTCI-based steady-state   calculations and the horizontal dashed line shows the reference value. The slight discrepancy in the height of the quasi-particle peak at low $\eta$ can be explained by the limited time window in the Fourier transform of the reference data. Because the conventional OCA implementation on the three-leg Kadanoff-Baym contour is numerically very costly, we restricted the calculation to $t_\text{max}=64$. In the more efficient QTCI-based steady-state approach, we instead use a time interval of length $t_\text{max} \approx 6000$. The latter time window is long enough that all the pp functions are fully decayed at relative time $t_\text{max}$, even for the smallest considered damping. 

	Generally, a larger $\eta$ enables a faster convergence of the DMFT self-consistency loop, at the cost of a broadened spectrum.
	We expect that this trend does not change at higher diagram orders $n$. While the effect of the $\eta$-damping can be substantial, and $\eta\lesssim 0.001$ should be used for low-temperature metallic systems, it is important to note that the related broadening effect is an intrinsic property of the employed pp based steady-state formalism, and not a consequence of the QTCI evaluation of the integrals. 

Looking at the convergence of the spectral function starting from the noninteracting semicircular density of states, we find that the first iteration generates two sharp peaks at roughly the expected position of the Hubbard bands, while subsequent iterations produce additional peaks, until a solution with smooth Hubbard bands emerges. The small wiggles in the Hubbard bands of the OCA solution 
could be a
remnant of these peaks, indicative of a not yet fully converged DMFT result, or they could originate from the inaccuracies of the QTCI fitting (we did not symmetrize the spectra in the DMFT calculation).

The computational cost of the QTCI-based OCA calculation is quite modest. One DMFT iteration for $2^{18}$ time points takes about 10 minutes on a single processor, where $\sim$10\% of the time is spent for the QTCI fitting, and most of the rest in the calculation of the frequency dependent pp self-energies and physical Green's function from the fitted $\sigma(v_1,v_2,v_3)$ and $g_\text{vert}(v_1,v_2,v_3)$, respectively.

\subsubsection{Nonequilibrium system}

\begin{figure}[t]
	\centering
	\includegraphics[width=1.0\columnwidth]{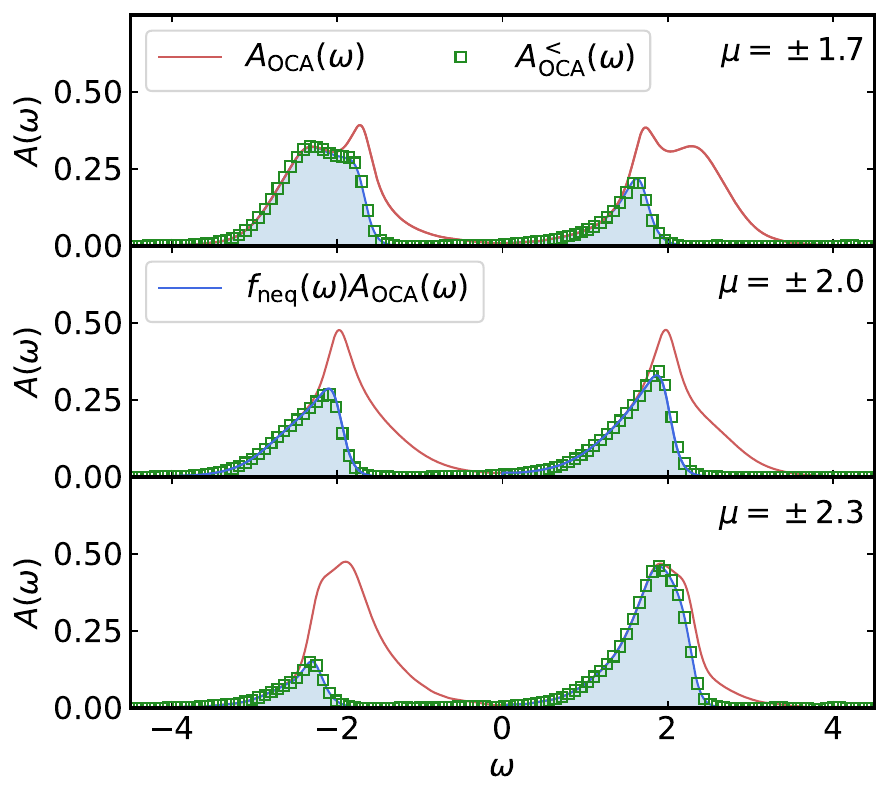}
	\caption{
		OCA spectral function of the photodoped Mott insulator with $U=4$ and $T_\text{eff}=0.1$.
		From top to bottom, the corresponding $|\mu_{\pm}|$ values are 1.7, 2.0, 2.3, and the damping factor is $\eta=0.001$.
		Red lines show the spectral function $A(\omega)$, blue lines represent the lesser (occupied) part imposed by multiplying $A(\omega)$ with $f_\text{neq}(\omega)$, while the green symbols show the measured occupation $A^<(\omega)$. 
	}
	\label{fig:photodoping}
\end{figure}

Finally, we show in Fig.~\ref{fig:photodoping} the nonequilibrium steady-state solutions of ``photodoped" Mott insulators with an excess density of doubly occupied and empty sites (doublons and holons). 
In previous nonequilibrium steady-state simulations of such systems, the photodoping was mimicked by weakly coupling full (empty) Fermion baths to the upper (lower) Hubbard bands \cite{Li2020,Li2021,Ray2023,Ray2024}. Recently, it was shown that a consistent description of such photodoped states can also be obtained by directly imposing the nonequilibrium distribution function $f_\text{neq}(\omega)$ \cite{Kuenzel2024}, which is easier to implement. 
Here, we follow the same strategy and define the lesser spectrum (occupied states) from the spectral function by $A^<(\omega)=f_\text{neq}(\omega) A(\omega)$, and similarly the greater spectrum (unoccupied states) by $A^>(\omega)=[1-f_\text{neq}(\omega)] A(\omega)$. Combined with the self-consistency equation (\ref{eq_selfconsistency}) and the relation between the correlation functions and spectral functions \cite{Aoki2014} one finds $\Delta^<(\omega)=-2i(\frac{D}{2})^2f_\text{neq}(\omega)  \text{Im}G^R(\omega)$, and similarly for $\Delta^>(\omega)$. 

The nonequilibrium distribution function is chosen as $f_\text{neq.}(\omega)=f_{T_\text{eff}}(\omega-\mu_\pm)$ for $\omega \gtrless 0$. The precise switching behavior near $\omega=0$ is not important, since we consider gapped systems. Here, $f_{T_\text{eff}}(\omega-\mu_\pm)$ is the Fermi function with the effective temperature $T_\text{eff}$ of the photodoped doublons and holons, and $\mu_\pm$ denotes the corresponding chemical potentials. The figure shows results for $U=4$, $T_\text{eff}=0.1$ and various $\mu_\pm$. The red spectrum is $A(\omega)$, the blue line $f_\text{neq}(\omega) A(\omega)$, while the green symbols show the measured $A^<(\omega)$. The good agreement between the latter two curves shows that the DMFT calculation converged to a self-consistent solution for the imposed effective temperature and chemical potentials. We found this to be generally the case for low photodopings and low enough $T_\text{eff}$, where the system remains fully gapped. If the photodoping is large and $T_\text{eff}$ is increased, the gap gets partially filled, which in a real system would lead to enhanced doublon-holon recombination and strong heating. In such a case, the adopted procedure no longer provides a consistent description of photodoped (quasi-)steady states. 

Near the doublon/holon Fermi levels $\mu_{\pm}$, there appear quasi-particle peaks in the nonequilibrium spectral function. For low photodoping, these peaks are associated with the doublon and holon distributions, which indicates that the mobile charge carriers are doublons and holons moving in the background of (immobile) half-filled states. For strong photodoping the peaks are instead associated with the remaining singlons, which become mobile in a half-filled system with a large density of doubly occupied and empty states. 

\begin{figure}[t]
	\centering
	\includegraphics[width=1.0\columnwidth]{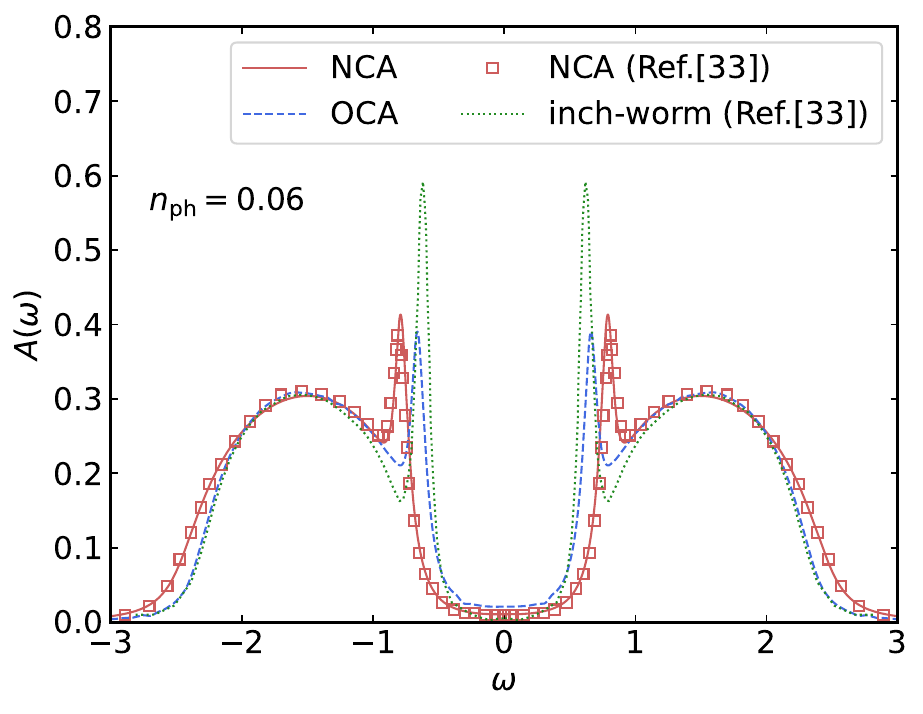}
	\caption{
		Comparision between NCA and OCA, and benchmark against the inchworm results from Ref.~\cite{Kuenzel2024} for the doublon density $n_{\text{ph}}=0.06$. 
		Here, the parameters are $T=T_{\text{eff}}=1/25\sqrt{2}, U=4/\sqrt{2}$, and $\eta=0.0001$.
		To enable a direct comparision with Ref.~\cite{Kuenzel2024}, the nonequilibrium distribution function $f_{\text{neq}}$ employs a smooth switching function near $\omega=0$. 
		The red boxes show the NCA spectrum from Ref.~\cite{Kuenzel2024}.
	}
	\label{fig:photodopingBenchmark}
\end{figure}
	
In Fig.~\ref{fig:photodopingBenchmark}, we also compare the OCA spectral function with the NCA one and the inchworm result reported in Ref.~\cite{Kuenzel2024}, which should be numerically exact upon full convergence in diagram orders. The parameters in this simulation are $T=T_{\text{eff}}=1/25\sqrt{2}, U=4/\sqrt{2}$, and $\eta=0.0001$. 
 Compared to the NCA result, the Mott gap shrinks and the quasi-particle peaks associated with the photodoped doublons and holons get more clearly separated from the Hubbard bands. Also the high-energy tail of the spectrum is shifted and approaches that of the inchworm spectrum.  
Although the trend in the gap-size is consistent with a convergence towards the inchworm spectrum with increasing order $n$, the discrepancy in the weight of the quasi-particle features remains remarkably large, given the moderate change from NCA to OCA.

\section{Conclusions}
\label{sec:conclusions}

We discussed the QTCI-based implementation of strong-coupling impurity solvers for equilibrium and nonequilibrium steady-state systems. Focusing on the OCA as the simplest nontrivial variant of such a solver, we showed how the pp self-energy $\Sigma(\omega)$ can be directly obtained in frequency space from the QTCI fit of the corresponding integrand $\sigma(v_1,v_2,v_3)$, which is a function of suitably defined time-difference variables $v_i$. The QTCI fit achieves a good accuracy of about four digits, even for the lesser component which involves discontinuities in the integrand. At least with the currently employed version of QTCI (the xfac code \cite{Fernandez2024}), however, a suitable choice of initial pivots is required. This limitation could be potentially overcome with more sophisticated global pivot search algorithms, which should be important for higher-order solvers. An alternative procedure would be to reparametrize the integral expression for the pp self-energy in such a way that no operators are shifted from the lower to the upper contour in the integrations. This simplifies the fitting, but comes at the cost of a larger number of QTCI fits.   

Our comparison to exactly solvable single bath-site results and conventional NCA and OCA implementations on the L-shaped Kadanoff-Baym contour showed that the damping factor $\eta$ used in the steady-state formalism can lead to some bias in the results, in particular a significant smearing of narrow quasi-particle peaks. To reduce this bias, a small $\eta$ factor and correspondingly large $t_\text{max}$ should be used. This difficulty is related to the slow decay of certain pp Green's functions which are needed for the calculation of the pp self-energy. In systems where the hybridization function decays fast, this problem is less severe, since the pp self-energy also involves such factors. The results for insulators and, importantly, the nonequilibrium steady state calculations for photodoped Mott systems converged rapidly with decreasing $\eta$. 

The $\eta$-broadening issue is intrinsic to the adopted pp formalism \cite{Li2021}, and not related to the QTCI implementation of the impurity solver. The pp specific challenges are absent in alternative solvers, like the influence functional approach \cite{Thoenniss2023}, which do not rely on pp Green's functions. 

Efficiency-wise, the TCI factorization leads to a significant speed-up compared to the conventional numerical quadrature scheme \cite{Eckstein2010}. For the presented DMFT calculations, the computational cost was reduced by roughly two orders of magnitude. The main bottleneck in the current implementation is the calculation of $\Sigma(\omega)$ from the fitted integrand $\sigma(v_1,v_2,v_3)$, which scales as $\mathcal{O}(N_\omega R (2n-1) N_\text{bond}^3)$ with $N_\omega$ the number of frequency points, $R$ the number of bits in the quantics representation of $v_i$, $n$ the order (here $n=2$) and $N_{\text{bond}}$ the bond dimension of the tensor train. Since we evaluate each frequency point separately, it would in principle be possible to employ a sparse adaptive grid, rather than the currently used equally spaced frequency grid. Such grids and additional optimizations in the implementation will further reduce the cost of the QTCI solver. 

The obvious next steps are the implementation of higher order solvers ($n\ge 3$) and the extension of our technique to electron-boson systems using the combined strong/weak-coupling approach \cite{Golez2015} and the Lang-Firsov scheme \cite{Werner2013}. Higher orders will be particularly relevant for multiorbital generalizations \cite{Strand2017}, where NCA and OCA cannot be expected to produce reliable results for metallic systems. It would also be interesting to compare the performance of the QTCI-based solver to the TCI strong coupling solver which was very recently presented in Ref.~\onlinecite{Eckstein2024}.  

\acknowledgements

We thank L. Geng, D. Golez, M. Eckstein and H. Shinaoka for helpful discussions. 
A.J.K. acknowledges support from DGIST Start-up Fund Program of the Ministry of Science and ICT (Grant No. 2024010026). 
We acknowledge the DGIST Supercomputing Big Data Center for the dedicated allocation of computing time.

\bibliography{ref}

\end{document}